\documentclass[showpacs,preprintnumbers,amsmath,amssymb]{revtex4}

\usepackage{graphicx}
\usepackage{dcolumn}
\usepackage{bm}
\begin{document}
\title{Nuclear effects in $F_3$ structure function of nucleon}
\author{M. Sajjad Athar$^1$, S. K. Singh$^1$ and M. J. Vicente Vacas$^2$}

\affiliation{$^1$Department of Physics, Aligarh Muslim University, Aligarh-202 002, India.\\
$^2$Departamento de F\'{\i}sica Te\'orica and IFIC, \\
Centro Mixto Universidad de
Valencia-CSIC,\\
46100 Burjassot (Valencia), Spain}
\date{\today}
\begin{abstract}
We study  nuclear effects in the $F^A_3(x)$ structure function  in the deep inelastic neutrino reactions on iron by using a relativistic framework to describe the nucleon spectral functions in the nucleus. 
 The results for the ratio $R(x,Q^2)=\frac{F^A_3(x,Q^2)}{AF^N_3(x, Q^2)}$ and the Gross-Llewellyn Smith(GLS) integral $G(x,Q^2)=\int_x^1 dx F^A_3(x,Q^2)$ in nuclei are discussed and compared with the recent results available in literature from theoretical and phenomenological analyses of experimental data.
\end{abstract}
\maketitle
\section{Introduction} 
The recent experimental results reported by the NuTeV collaboration~\cite{Tzanov} on  weak charged and neutral current induced (anti)neutrino processes on an iron target in the deep inelastic region have emphasized the importance of nuclear medium effects. 
 There are many theoretical analyses of the deep inelastic scattering of charged leptons from
nuclear targets where various nuclear effects like shadowing, anti-shadowing,  Fermi motion and binding of the nucleons 
in various kinematic regions have been studied. These have been discussed in several review articles~\cite{Armesto, Petti, Geesaman}, but in the case of deep inelastic scattering of (anti)neutrinos from nuclear targets, there are  few 
calculations where the dynamical origin of the nuclear medium effects has been studied~\cite{Petti1, Kulagin, Singh, Qiu, Kovalenko}. In some theoretical analyses, nuclear medium effects have been phenomenologically described in terms of a few parameters which are determined from fitting the experimental data of charged leptons and (anti)neutrino deep inelastic scattering from various nuclear targets~\cite{Hirai, Bodek1, Eskola, Schienbein, Florian}. 

The differential scattering cross section for the deep inelastic scattering of (anti)neutrinos from unpolarized nucleons in the limit of lepton mass $m_l \rightarrow 0$, is described in terms of three structure functions, $F^\nu_1$($x$,$Q^2$), $F^\nu_2$($x$,$Q^2$) and $F^\nu_3$($x$,$Q^2$), where $x=\frac{Q^2}{2M\nu}=-\frac{q^2}{2M\nu}$ is the Bjorken variable, $\nu$ and $q$ being the energy and momentum transfer of leptons. In the asymptotic region of Bjorken scaling i.e. $Q^2 \rightarrow \infty$, $\nu \rightarrow \infty$, $x$ finite, all the structure functions depend only on the Bjorken variable $x$. In this scaling limit, $F^\nu_1(x)$ and $F^\nu_2(x)$ are related by the Callan-Gross relation~\cite{Callan} leading to only two independent structure functions $F^\nu_2$($x$) and $F^\nu_3$($x$) which are determined from the experimental data on deep inelastic scattering of (anti)neutrinos in the asymptotic region. Scaling violation effects have been studied in deep inelastic scattering of (anti)neutrino from nucleon targets using methods of perturbative and non-perturbative QCD. The $Q^2$ dependence of the structure functions is determined from the $Q^2$ evolution given by Dokshitzer-Gribov-Lipatov-Altarelli-Parisi (DGLAP) equations~\cite{DGLAP} obtained in perturbative QCD calculations. Non perturbative $Q^2$ corrections to the structure functions have been also studied by many authors~\cite{Sidorov, Dasgupta}. 

Since most of the experimental data on deep inelastic scattering of high energy (anti)neutrinos are obtained on nuclear targets over a wide range of $Q^2$, it is important to study the nuclear medium modification effects on these structure functions $F^\nu_2$($x$,$Q^2$) and $F^\nu_3$($x$,$Q^2$) specially in iron where most recent, high statistics, high precision data are available~\cite{Tzanov}. 
The nuclear medium modification effects on the average structure function $F^A_3$($x$,$Q^2$)=$\frac{1}{2}$($F^{\nu A}_3$($x$,$Q^2$) + $F^{\bar\nu A}_3$($x$,$Q^2$)) on isoscalar nuclear targets arise mainly due to modification of the valence quark parton distributions in the nuclear medium. The contributions of sea quarks and gluons in a quark parton picture vanish in the case of a symmetric sea and arise only when the various sea quark parton distribution are taken to be different. 
 Of much interest is the  estimate of the size of the nuclear medium effects on the GLS sum rule~\cite{GLS} which is 
measured over a wide range of $Q^2$ for deep inelastic scattering of (anti)neutrinos on nuclear targets and 
has been used to determine  the QCD coupling constant $\alpha_s$ ~\cite{Kim}.
 
In this paper, we  study some nuclear medium  effects on the nucleon structure function $F^A_3$($x$,$Q^2$) in iron. We use a theoretical spectral function to describe the momentum distribution of nucleons in the nucleus. The spectral function has been calculated using the Lehmann's representation for the relativistic nucleon propagator and nuclear many body theory is used to calculate it for an interacting Fermi sea in nuclear matter. A local density approximation is then applied to translate these results to finite nuclei. The method has been earlier used successfully to describe the EMC and other nuclear effects in deep inelastic scattering of charged leptons from nuclei~\cite{Marco}. Our approach describes an alternative method to calculate nuclear medium effects in the nucleon structure functions using  nuclear many body theory different from the approach used by Kulagin~\cite{Kulagin}. Our model has certain limitations as it only attempts to consider the modifications of nucleonic contributions to $F^N_3$($x$,$Q^2$) arising due to binding energy, off mass shell and Fermi motion  of the nucleon in the nuclear medium which dominate in the region of $x\ge$0.3. In the region of 0.3$>x>$0.1, corresponding to the anti-shadowing region, the nuclear medium modification effects on $F^A_3$($x$,$Q^2$) are expected to be small due to vanishing of the pion contribution in this model unlike $F^{A l}_2$($x$,$Q^2$) where contributions of pion and rho mesons play a dominant role~\cite{Marco}. We do not consider the shadowing region of 0.0$<x<$0.1 in this paper and that will be considered in the extension of this work. Therefore, the results presented here should be able to describe the dominant contribution of nuclear medium effects to $F^A_3$($x$,$Q^2$) in the range of 0.1$<x<$1.

In sections 2 and 3, we describe briefly the formalism used to calculate nuclear effects and present our results and discussion in section 4.
\section{Deep Inelastic Neutrino Nucleon Scattering}
The cross section for the reaction
\begin{equation} 	\label{reaction}
\nu_l(\bar\nu_l) + N \rightarrow l^-(l^+) + X,
\end{equation}
 is given by:
\begin{eqnarray} \label{csection}
\sigma&=&\frac{1}{v_{rel}}\frac{2m_\nu}{2E_{\nu}({\bf k})}\frac{2M}{2E({\bf p})}\int\frac{d^3 k^{\prime}}{2\pi^{3}}\frac{2m_l}{2E_{l}({\bf k}^{\prime})}\\ \nonumber
&&\times \prod_{i=1}^{N}\int\frac{d^3 p_{i}^{\prime}}{2\pi^{3}}\prod_{l\epsilon f}\left(\frac{2M^{\prime}_{l}}{2E^{\prime}_{l}}\right)\prod_{j\epsilon b}\left(\frac{1}{2\omega^{\prime}_{j}}\right)\bar{\sum} \sum \left|T\right|^{2}(2\pi)^4\\ \nonumber
&&\times \delta^{4}\left(p+k-k^{\prime}-\sum_{i=1}^{N}p_{i}^{\prime}\right)
\end{eqnarray}
where $f$ stands for
fermions and $b$ for bosons in
the final state $X$. The index $i$ is split
into $l$ and $j$ for fermions and bosons respectively,

$T$ is the invariant matrix element for the above reaction and is,
written as
\begin{equation}	\label{Tinv}
- i T = \left( \frac{i G}{\sqrt{2}}\right) 
\bar{u}_l (k^\prime) 
\gamma^{\alpha} (1 -\gamma_5)u_{l} (k) \,
\left(\frac{m_W^2}{q^2-m_W^2}\right)\langle X | J_{\alpha} | N \rangle\,.
\end{equation}

After performing the phase space integration in Eq.(\ref{csection}), the double differential scattering cross section evaluated for a nucleon target in its rest frame is expressed as,
\begin{equation} 	\label{dif_cross}
\frac{d^2 \sigma_{\nu,\bar\nu}^N}{d \Omega' d E'} 
= \frac{{G_F}^2}{(2\pi)^2} \; \frac{|{\bf k}^\prime|}{|{\bf k}|} \;
\left(\frac{m_W^2}{q^2-m_W^2}\right)^2
L^{\alpha \beta}_{\nu, \bar\nu}
\; W_{\alpha \beta}^{N}\,,
\end{equation}
where $G_F$ is the Fermi coupling constant, $m_W$ is the mass of the W boson, $l(=e,\,\mu)$ is a lepton, $k$ is the incoming  neutrino four momentum  and $k^\prime$ is the outgoing four momentum of the lepton, $q=k-k^\prime$ is the four momentum transfer and $\Omega', E'$ refer to the outgoing lepton. $N$ is a nucleon, $X$ is a jet of n hadrons consisting of fermions(f) and bosons(b) in the final state labeled by $l$ and $j$ in the following.

The lepton tensor for antineutrino(neutrino) scattering $L^{\alpha \beta}$ is given by
\begin{equation} 	\label{dif_cross2}
L^{\alpha \beta}=k^{\alpha}k'^{\beta}+k^{\beta}k'^{\alpha}
-k.k^\prime g^{\alpha \beta} \pm i \epsilon^{\alpha \beta \rho \sigma} k_{\rho} 
k'_{\sigma}\,,
\end{equation}
and the hadronic tensor $W^{N}_{\alpha \beta}$ is defined as
\begin{eqnarray}\label{hadten}
W^{N}_{\alpha \beta} = \frac{1}{2\pi} \bar{\sum_{s_N}} \;
\sum_X \; \sum_{s_i} \prod^n_{i = 1}
\; \int \frac{d^3 p'_i}{(2 \pi)^3} 
\prod_{l \epsilon f} \;
\left(
\frac{2 M'_l}{2 E'_l}
\right) \;
\prod_{j \epsilon b} \;
\left(
\frac{1}{2 \omega'_j}
\right)
\langle X | J_{\alpha} | N \rangle 
\langle X | J_{\beta} | N \rangle^* 
(2 \pi)^4 \delta^4 (p + q - 
\sum^n_{i = 1} p'_i)\,,
\end{eqnarray}
where $q$ is the momentum of the virtual $W$, $s_N$ the spin of the
nucleon and $s_i$ the spin of the fermions in $X$. In the case of antineutrino $\langle X | J_{\alpha} | N \rangle$ is replaced by $\langle X | J^{\dagger}_{\alpha} | N \rangle$.

The most general form of the hadronic tensor $W^{N}_{\alpha \beta}$ is expressed as~\cite{Bilenky}:
\begin{eqnarray} \label{had_ten}
W^{N}_{\alpha \beta} =&& 
\left( \frac{q_{\alpha} q_{\beta}}{q^2} - g_{\alpha \beta} \right) \;
W_1^{\nu (\bar{\nu})}
+ \frac{1}{M^2}\left( p_{\alpha} - \frac{p . q}{q^2} \; q_{\alpha} \right)
\left( p_{\beta} - \frac{p . q}{q^2} \; q_{\beta} \right)
W_2^{\nu (\bar{\nu})}-\frac{i}{2M^2} \epsilon_{\alpha \beta \rho \sigma} p^{\rho} q^{\sigma}
W_3^{\nu (\bar{\nu})} + \nonumber\\
&&
\frac{1}{M^2} q_{\alpha} q_{\beta}
W_4^{\nu (\bar{\nu})}
+\frac{1}{M^2} (p_{\alpha} q_{\beta} + q_{\alpha} p_{\beta})
W_5^{\nu (\bar{\nu})}
+ \frac{i}{M^2} (p_{\alpha} q_{\beta} - q_{\alpha} p_{\beta})
W_6^{\nu (\bar{\nu})}\,,
\end{eqnarray}
where $M$ is the nucleon mass and $W_i^{N}$ are the structure functions, which depend
on the scalars $q^2$ and $p.q$. The terms depending on $W_4$, $W_5$
and $W_6$ in Eq.~(\ref{had_ten}) do not contribute to the cross
section in Eq. (\ref{dif_cross}) in the limit of lepton mass $m_l \rightarrow 0$.

In terms of the Bjorken variables $x$ and y defined as
\begin{eqnarray}	\label{Bj_var}
x=\frac{Q^2}{2M\nu}, \quad y=\frac{\nu}{E_{\nu}}, \quad Q^2=-q^2, \quad \nu=\frac{p.q}{M}
\end{eqnarray}
we can write the expression for the differential scattering cross section (in the limit of
lepton mass $m_l \rightarrow 0$) as

\begin{eqnarray}\label{diff_dxdy}
\frac{d^2 \sigma^{\nu(\bar{\nu})}}{d x d y}&=&
\frac{{G_F}^2ME_{\nu}}{\pi}
\left\{xy^2 F_1^{\nu(\bar{\nu})} (x, Q^2)
+ \left(1-y-\frac{xyM}{2 E_{\nu}}\right) F_2^{\nu(\bar{\nu})} (x, Q^2)
\pm  xy(1-y/2)F_3^{\nu(\bar{\nu})} (x, Q^2)
\right\}\,,
\end{eqnarray}
where the $+$ ($-$) sign stands for the neutrino (antineutrino)
cross section, and the $F_i^{\nu, \bar\nu} (x,Q^2)$ are dimensionless
structure functions defined as 
\begin{eqnarray}\label{relation}
F_1^{\nu(\bar{\nu})}(x, Q^2)&=&M W_1^{\nu(\bar{\nu})}(\nu, Q^2) \\ \nonumber
F_2^{\nu(\bar{\nu})}(x, Q^2)&=&\nu W_2^{\nu(\bar{\nu})}(\nu, Q^2) \\ \nonumber
F_3^{\nu(\bar{\nu})}(x, Q^2)&=&\nu W_3^{\nu(\bar{\nu})}(\nu, Q^2).
\end{eqnarray}
In the Bjorken limit of scaling valid in the asymptotic region i.e. $Q^2 \rightarrow \infty$, $\nu \rightarrow \infty$, $x$ finite, the structure functions $F_i^{\nu(\bar{\nu})}(x, Q^2)$ are independent of $Q^2$ and depend only on the single dimensionless variable $x$, and satisfy the Callan-Gross relation~\cite{Callan} given as $2xF_1(x)=F_2(x)$. Using this relation, the cross section in Eq.(\ref{diff_dxdy}) is described in terms of two independent structure functions $F_2(x)$ and $F_3(x)$. In the quark parton model of deep inelastic scattering, in the Bjorken scaling limit, these structure functions are determined in terms of parton distribution functions for quarks and antiquarks.

Specifically, the structure function $F_3(x)$ is given as:

$$
F^{\nu p}_{3}= 2[d(x) + s(x) -{\bar u}(x) - {\bar c}(x)]\,,
$$
$$
F^{\nu n}_{3}= 2[u(x) + s(x) -{\bar d}(x) - {\bar c}(x)]\,,
$$
$$
F^{{\bar \nu} p}_{3}= 2[u(x) + c(x) -{\bar d}(x) - {\bar s}(x)]\,,
$$
$$
F^{{\bar \nu} n}_{3}= 2[d(x) + c(x) -{\bar u}(x) - {\bar s}(x)]\,,
$$
 The average structure function $F_{3}^{N}(x)$ on isoscalar nucleon target defined as
\begin{eqnarray*}\nonumber
F_{3}^{N}(x)=\frac{1}{2}\left(F^{\nu N}_{3} + F^{\bar\nu N}_{3}\right)
\end{eqnarray*} 
is given by
\begin{eqnarray*}\nonumber
F_{3}^{N}(x)=[ u_v(x) + d_v(x) + s(x)- {\bar s}(x) + c(x) - {\bar c}(x)],
\end{eqnarray*}
where $u_v(x)=u(x) - {\bar u}(x)$ and $d_v(x)=d(x) - {\bar d}(x)$ are the valence quark parton distributions.

Thus, for an isoscalar target and a symmetric sea, $F_{3}^{N}(x)$ structure function is given in terms of valence quarks $u_v$ and $d_v$ which satisfies the Gross-Llewellyn Smith sum rule~\cite{GLS}:
\begin{equation}	\label{gls}
\int_0^1 F^N_3(x)dx=3.
\end{equation}
In the non-asymptotic region, scaling violations occur and the structure function $F_i(x)$ acquire $Q^2$ dependence which are calculated with the DGLAP equations of $Q^2$ evolution obtained using perturbative QCD. As a consequence, the Callan Gross relation and the Gross-Llewellyn Smith sum rule are modified as follows:
\begin{equation}	\label{modcgr}
2xF_1(x,Q^2)=F_2(x,Q^2)\frac{1+\frac{4M^2x^2}{Q^2}}{1+R(x,Q^2)},
\end{equation}
where R=$\frac{\sigma_L}{\sigma_T}$ is the ratio of the cross section of longitudinally to transversely polarized W bosons~\cite{Whitlow} and
\begin{eqnarray}\label{modgls}
S(Q^2)&=&\int_0^1 F_3(x,Q^2)dx=3\left(1-\frac{\alpha_s}{\pi}-a(n_f)\left(\frac{\alpha_s}{\pi}\right)^2-b(n_f)\left(\frac{\alpha_s}{\pi}\right)^3\right)-\frac{\Delta_{HT}}{Q^2},
\end{eqnarray}
where $a(n_f)$ and $b(n_f)$ are functions of the number of quark flavors accessible at a given $Q^2$ and $\Delta_{HT}$ is the higher twist correction~\cite{Larin}.
\begin{figure}
\includegraphics{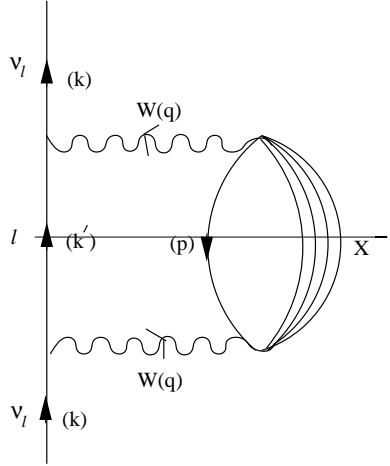}
\caption{Self-energy diagram of the neutrino in the nuclear medium
associated with the process of deep inelastic neutrino-nucleon
scattering. The imaginary part is calculated by cutting along the horizontal line and applying the Cutkosky rule for putting the particles on the mass shell.}
\end{figure}
\section{Nuclear effects in neutrino scattering}
When the reaction given by Eq.(\ref{reaction}) takes place on a nucleon in the nucleus, nuclear effects have to be considered. There are two main nuclear effects. Firstly, a kinematic effect which arises as the struck nucleon is not at rest but is moving with a Fermi momentum in the rest frame of the nucleus, leading to a Lorentz contraction of the incident flux used in deriving Eq.(\ref{dif_cross}). Secondly, the more important dynamic effects which arise due to Fermi motion, Pauli blocking and strong interaction of the initial nucleon in the nuclear medium. 

In a nuclear medium the expression for the cross section given in Eq.(\ref{dif_cross}) is modified as:
\begin{equation} 	\label{dif_cross_nucleus}
\frac{d^2 \sigma_{\nu,\bar\nu}^A}{d \Omega' d E'} 
= \frac{G_F^2}{(2\pi)^2} \; \frac{|{\bf k}^\prime|}{|{\bf k}|} \;
\left(\frac{m_W^2}{q^2-m_W^2}\right)^2
L^{\alpha \beta}_{\nu, \bar\nu}
\; W_{\alpha \beta}^{A}\,,
\end{equation}
where $W_{\alpha \beta}^{A}$ is the nuclear hadronic tensor defined in terms of nuclear hadronic structure functions $W_{i A}(x,Q^2)$ through Eq.(\ref{had_ten}). Accordingly, the dimensionless structure functions $F^A_{i}(x,Q^2)$ are defined in terms of  $W^A_i(x,Q^2)$ through Eq.(\ref{relation}). In our formalism the neutrino nuclear cross sections are obtained in terms of neutrino self energy $\Sigma(k)$ in the nuclear medium which also defines the dimensionless nuclear structure functions $F^A_{i}(x,Q^2)$. A comparison with Eq.(\ref{dif_cross_nucleus}) gives the nuclear hadronic tensor $W_{\alpha \beta}^{A}$. With proper choice of tensor components $\alpha,\beta$, the dimensionless nuclear structure functions $F^A_{i}(x,Q^2)$ are obtained\cite{Marco}.

The neutrino self-energy in nuclear matter corresponding to Fig.1 is given by,
\begin{eqnarray}\label{Sigma}
\Sigma (k) =&& (-i)\frac{G_F}{\sqrt{2}}\frac{4}{m_\nu}
\int \frac{d^4 k^\prime}{(2 \pi)^4} \frac{1}{{k^\prime}^2-m_l^2+i\epsilon} 
\left(\frac{m_W }{q^2-m_W^2}\right)^2 \; L_{\alpha \beta} ~~ \Pi^{\alpha \beta} (q)\,,
\end{eqnarray}
where $L_{\alpha\beta}$ is given by Eq.(\ref{dif_cross2}) and $\Pi^{\alpha \beta} (q)$ is the $W$ self-energy in the nuclear medium and is written with the help of Fig.2 as:
\begin{eqnarray}\label{Self_1}
-i\Pi^{\alpha \beta} (q) =&&
(-) \; \int \frac{d^4 p}{(2 \pi)^4} iG(p) \;  \;
\sum_X \; \sum_{s_p, s_i} \prod^n_{i = 1}
\int \frac{d^4 p'_i}{(2 \pi)^4}
\prod_l i G_l (p'_l) \prod_j \; i D_j (p'_j) 
\left( \frac{-G_F m_W^2}{\sqrt{2}} \right)\nonumber\\
&&\times
\langle X | J^{\alpha} | N \rangle \langle X | J^{\beta} | N \rangle^*
(2 \pi)^4 \delta^4 (q + p - \Sigma^n_{i = 1} p'_i)\,.
\end{eqnarray}
In the above expression  $G_l (p'_l)$ and $D_j (p'_j)$ are respectively the nucleon and meson relativistic propagators in the final state which are taken as the standard free relativistic propagators~\cite{Itzykson}. $G(p)$ is the  nucleon propagator with mass M and energy $E({\bf p})$ in the initial state, which is calculated for a relativistic nucleon in the interacting Fermi sea.

For this we start with the relativistic Dirac propagator G(p) for a free nucleon, which is written in terms of the contribution from the positive and negative energy components of the nucleon described by the Dirac spinors $u({\bf p})$ and $v({\bf p})$ using their appropriate normalisations~\cite{Itzykson} as
\begin{equation} \label{prop1}
G^{0}(p)=\frac{\not p+M}{p^2-M^2+i\epsilon}=\frac{M}{E(p)}\left\{\frac{\sum_{r}u_{r}({\bf p})\bar u_{r}({\bf p})}{p^{0}-E({\bf p})+i\epsilon}+\frac{\sum_{r}v_{r}(-{\bf p})\bar v_{r}(-{\bf p})}{p^{0}+E({\bf p})-i\epsilon}\right\}
\end{equation}

The  nucleon propagator G(p) is then calculated by making a perturbative expansion of $G(p)$ in terms of $G^{0}(p)$ given in Eq.(\ref{prop1}) by retaining the positive energy contributions only(the negative energy components are suppressed).
This perturbative expansion is summed in ladder approximation to give~\cite{Marco}:
\begin{eqnarray}
G(p)&=&\frac{M}{E({\bf p})}\sum_{r}u_{r}({\bf p})\bar u_{r}({\bf p})\frac{1}{p^{0}-E({\bf p})}\\ \nonumber
&&+\frac{M}{E({\bf p})}\sum_{r}\frac{u_{r}({\bf p})\bar u_{r}({\bf p})}{p^{0}-E({\bf p})}\sum(p^{0},{\bf p})\frac{M}{E({\bf p})}\sum_{s}\frac{u_{s}({\bf p})\bar u_{s}({\bf p})}{p^{0}-E(p)}+.....\\ \nonumber
&&=\frac{M}{E({\bf p})}\sum_{r}\frac{u_{r}({\bf p})\bar u_{r}({\bf p})}{p^{0}-E({\bf p})-\bar u_{r}({\bf p})\sum^N(p^{0},{\bf p})u_{r}({\bf p})\frac{M}{{\bf E(p)}}}
\end{eqnarray}
where $\Sigma^N(p^0,p)$ is the nucleon self energy in nuclear matter taken from Ref.~\cite{Fernandez}.
The relativistic nucleon propagator G(p) in a nuclear medium is then expressed as~\cite{Marco}:
\begin{eqnarray}\label{Gp}
G (p) =&& \frac{M}{E({\bf p})} 
\sum_r u_r ({\bf p}) \bar{u}_r({\bf p})
\left[\int^{\mu}_{- \infty} d \, \omega 
\frac{S_h (\omega, \bbox{p})}{p^0 - \omega - i \eta}
+ \int^{\infty}_{\mu} d \, \omega 
\frac{S_p (\omega, \bbox{p})}{p^0 - \omega + i \eta}\right]\,,
\end{eqnarray}
where $S_h (\omega, \bbox{p})$ and $S_p (\omega, \bbox{p})$ being the hole
and particle spectral functions respectively, which are given by~\cite{Marco,Fernandez}:
\begin{equation}\label{sh}
S_h (\omega, {\bf p})=\frac{1}{\pi}\frac{\frac{M}{E({\bf p})}Im\Sigma^N(p^0,p)}{(p^0-E({\bf p})-\frac{M}{E({\bf p})}Re\Sigma^N(p^0,p))^2
+ (\frac{M}{E({\bf p})}Im\Sigma^N(p^0,p))^2}
\end{equation}
for $p^0 \le \mu$
\begin{equation}
S_p (\omega, {\bf p})=-\frac{1}{\pi}\frac{\frac{M}{E({\bf p})}Im\Sigma^N(p^0,p)}{(p^0-E({\bf p})-\frac{M}{E({\bf p})}Re\Sigma^N(p^0,p))^2
+ (\frac{M}{E({\bf p})}Im\Sigma^N(p^0,p))^2}
\end{equation}
for $p^0 > \mu$. 

The normalisation of this spectral function is obtained by imposing the baryon number conservation following the method of 
Frankfurt and Strikman~\cite{Frankfurt}. For this purpose we evaluate the electromagnetic form factor at q=0, assuming baryons 
have unit charge for the purpose of normalisation, corresponding to Fig.2a, i.e.
\begin{eqnarray}\label{NNmu}
\left<N|B^{\mu}|N\right>\equiv \bar u({\bbox p}) \gamma^{\mu} u({\bbox p})~=~B \frac{p^{\mu}}{M};~ B=1,~ p^{\mu}\equiv(E({\bf p}),{\bf p})
\end{eqnarray}

\begin{figure}
\includegraphics{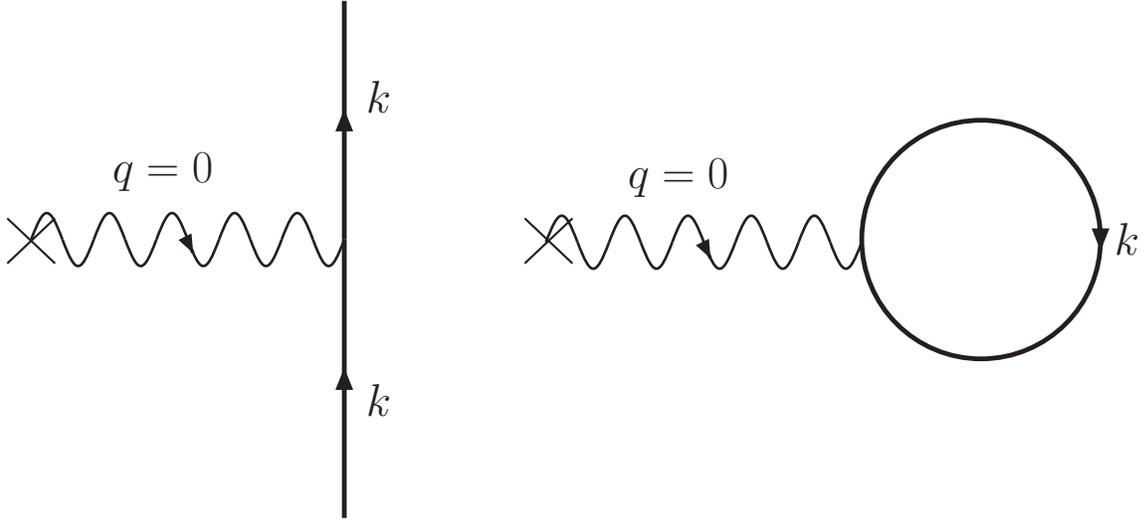}
\caption{Electromagnetic form factors for the case (a) free nucleon, (b) Fermi sea with B baryons.}
\end{figure}

When the nucleons are in the nuclear medium the many-body diagram shown by Fig.2b is evaluated~\cite{Marco} and gives
\begin{eqnarray}\label{AAmu}
\left<A|B^{\mu}|A\right>=-\int \frac{d^4p}{(2\pi)^4}V i Tr[G(p^0,{\bbox p})\gamma^\mu]e^{ip^0\eta}.
\end{eqnarray}
where V is the volume of the normalisation box and $exp({ip^0\eta})$, with $\eta \rightarrow 0^+$, is the convergence 
factor for loops appearing at equal times.

Using the expression given by Eq.(\ref{Gp}), it can be seen that the convergence factor limits the contribution of the hole spectral function and this gives
\begin{eqnarray} \label{ABmu}
\left<A|B^{\mu}|A\right>&=&V\int\frac{d^{3}p}{(2\pi)^3}\frac{M}{E({\bf p})}{Tr} \left[\sum_{r}u_{r}({\bf p})\bar u_{r}({\bf p})\gamma^{\mu}\right]\int_{-\infty}^{\mu}S_{h}(\omega,p)d\omega \nonumber\\
&&=V\int\frac{d^{3}p}{(2\pi)^3}\frac{M}{E({\bf p})}{Tr} \left[\frac{(\not p+M)_{on shell}}{2M}\gamma^{\mu}\right]\int_{-\infty}^{\mu}S_{h}(\omega,p)d\omega \nonumber\\
&&=2V\int\frac{d^{3}p}{(2\pi)^3}\frac{M}{E({\bf p})}\frac{p^{\mu}_{on shell}}{M}\int_{-\infty}^{\mu}S_{h}(\omega,p)d\omega\equiv
B\frac{P_A^\mu}{M_A} .
\end{eqnarray}
It is to be noted that in the last step we have imposed that this matrix element gives the right current with B baryons, 
in analogy to the expression given by Eq.~(\ref{NNmu}). $P_A^\mu$ is the four momentum of the nucleus. The operator 
$(\not p + M)$ comes from $\sum u_r({\bf p})\bar u_r({\bf p})$ which depends only on ${\bf p}$, and that corresponds to a 
free particle with $(p^\mu=E({\bf p}),{\bf p})$, therefore, the operator $(\not p + M)$ is written on shell. 
The last step in Eq.~(\ref{ABmu})) is obtained after performing the trace. Evaluating Eq.~(\ref{ABmu}) in the rest frame
 of the nucleus we obtain the normalisation as
\begin{eqnarray} \label{norm1}
2V\int\frac{d^{3}p}{(2\pi)^{3}}\int_{-\infty}^{\mu}S_{h}(\omega,p) d\omega= B~=1.
\end{eqnarray}
In the local density approximation, we do not have a box of constant density, and the reaction takes place at a 
point ${\bf r}$, lying inside a volume element $d^3r$ with local density $\rho_{p}({\bf r})$ and $\rho_{n}({\bf r})$ 
corresponding to the proton and neutron. Therefore, the upper limit in the integration over nucleon momentum in Eq.~(\ref{norm1}) is the local Fermi momentum $k_{F_{p,n}}({\bf r})$ of the nucleon given by:
 \begin{eqnarray} \label{Fermi1}
k_{F_p}({\bf r})= \left[ 3\pi^{2} \rho_{p}({\bf r})\right]^{1/3}, k_{F_n}({\bf r})= \left[ 3\pi^{2} \rho_{n}({\bf
r})\right]^{1/3}.
\end{eqnarray}
This makes the spectral function $S_{h}(\omega,p)$ density dependent i.e. $S_{h}(\omega,p,k_{F}({\bf r}))$
 and the normalisation condition given in Eq.~(\ref{norm1}) is modified to
\begin{eqnarray} \label{norm2}
2\int\frac{d^{3}p}{(2\pi)^{3}}\int_{-\infty}^{\mu}S_{h}(\omega,p,k_{F_{p,n}}({\bf r})) d\omega= \rho_{p,n}({\bf r})
\end{eqnarray}
For a symmetric nuclear matter of density $\rho({\bf r})$, there is a unique Fermi momentum given by $k_{F}({\bf r})= \left[\frac{3}{2}\pi^{2} \rho({\bf r})\right]^{1/3}$ for which we obtain
\begin{eqnarray} \label{norm3}
4\int\frac{d^{3}p}{(2\pi)^{3}}\int_{-\infty}^{\mu}S_{h}(\omega,p,k_{F}({\bf r})) d\omega= \rho({\bf r}),
\end{eqnarray}
where $\rho(r)$ is the baryon density for the nucleus.

This leads to the normalisation condition~\cite{footnote} given by
\begin{equation}\label{norm4}
4 \int d^3 r \;  \int \frac{d^3 p}{(2 \pi)^3} 
\int^{\mu}_{- \infty} \; S_h (\omega, \bbox{p}, \rho(r)) 
\; d \omega = A
\end{equation}

In the antineutrino case the expressions obtained are very similar.
$L_5^{\alpha \beta}$ appears, as in Eq.~(\ref{dif_cross2}), with
a minus sign in front and in the $W$ self-energy, Eq.~(\ref{Self_1}),
we have $\langle X | J^{\dagger}_{\alpha} | N \rangle$, instead
of $\langle X | J_{\alpha} | N \rangle$.

The probability per unit time for the neutrino to collide with nucleons
when traveling through nuclear matter is:
\begin{equation}
\Gamma (k) = - \frac{2 m_{\nu}}{E_{\nu} (\bbox{k})} \; 
\mbox{Im} \; \Sigma (k)\,,
\end{equation}
and the cross section for neutrino scattering from an element of volume $d^3 r$ and surface dS in the nucleus is given by
\begin{eqnarray}\label{change}
d \sigma & = & \Gamma d t d S = \Gamma \frac{dt}{dl} \;
dl dS = \frac{\Gamma}{v} \; d^3 r = \Gamma \; \frac{E_{\nu} (\bbox{k})}{|\bbox{k}|}
 \; d^3 r = - \frac{2 m_{\nu}}{|\bbox{k}|} \;
\hbox{Im} \; \Sigma \; d^3 r \,.
\end{eqnarray}
Using Eq.(\ref{change}) in Eq.(\ref{Sigma}), we get the expression for the total scattering cross section in the local 
density approximation as
\begin{eqnarray}\label{Sigma1}
\sigma =\frac{4\sqrt{2} G_F}{|\bf k|} \hbox{Im}
\int d^3r \int \frac{d^4 k^\prime}{(2 \pi)^4} \frac{1}{k^{\prime 2}-m_\mu^2+i\epsilon}
\left(\frac{m_W }{q^2-m_W^2}\right)^2 \; L_{\alpha \beta} \Pi^{\alpha \beta} (q)\,.
\end{eqnarray}
The imaginary part of the neutrino self energy in Eq.(\ref{Sigma1}) is evaluated by means of the Cutkosky rules 
\cite{Itzykson} by cutting the Feynman diagram shown in Fig.~3 along the dotted line which puts the particles corresponding 
to the cut propagators on the mass shell by replacing the fermion and meson propagators by their imaginary parts as
\begin{eqnarray}	
\label{Cutkosky}
\Sigma(k) & \rightarrow & 2 i Im \Sigma(k) \nonumber \\
D({p_j}^\prime) & \rightarrow & 2 i \theta(p_{0j}) ~Im D({p_j}^\prime)\nonumber \\
G({p_l}^\prime) & \rightarrow & 2 i \theta({p_{0l}}^\prime) ~ Im G({p_l}^\prime)\nonumber \\
\frac{1}{k^{\prime 2}-m_l^2+i\epsilon}& \rightarrow & 2 \pi \delta(k^{\prime 2}-m_l^2).
\end{eqnarray}
After performing the ${k_0}^\prime$, ${p_0}^\prime$ and $p_0$ integrations for all momenta in Eq.(\ref{Sigma1}) using Eqs.({\ref{Gp}) and ({\ref{Cutkosky}), we get the differential scattering cross section which is written in the local density approximation as:  
\begin{equation} 	\label{cross_nuclear}
\frac{d^2 \sigma_{\nu,\bar\nu}^A}{d \Omega' d E'} 
= \frac{{G_F}^2}{(2\pi)^2} \; \frac{|{\bf k}^\prime|}{|{\bf k}|} \;
\left(\frac{m_W^2}{q^2-m_W^2}\right)^2
L^{\alpha \beta}_{\nu, \bar\nu}
\; W_{\alpha \beta}^{A}\,,
\end{equation}
where
\begin{equation}	\label{conv_WA}
W^A_{\alpha \beta} = 4 \int \, d^3 r \, \int \frac{d^3 p}{(2 \pi)^3} \, 
 \int^{\mu}_{- \infty} d p^0 \frac{M}{E ({\bf p})} S_h (p^0, \bbox{p}, \rho(r))
W^N_{\alpha \beta} (p, q), \,.
\end{equation}
with $W^N_{\alpha \beta} (p, q)$ is given by Eq.(\ref{hadten}). 

Note that the factor $\frac{M}{E ({\bf p})}$ in Eq.~(\ref{conv_WA}) comes naturally in our formalism, when we perform 
the various momentum integrations in Eq.~(\ref{Sigma1}) to calculate the imaginary part using Eqs.~(\ref{Gp}) and 
(\ref{Cutkosky}). This can be physically understood as a kinematic factor which appears in the cross section defined in 
Eq.~(\ref{csection}) for a nucleon moving with momentum $p=(E,{\bf p})$ in the rest frame of the nucleus 
leading to the equation:
\begin{equation} 	\label{dif_cross_nucleus}
\frac{d^2 \sigma_{\nu,\bar\nu}^A}{d \Omega' d E'} 
= \frac{G_F^2}{(2\pi)^2} \; \frac{|{\bf k}^\prime|}{|{\bf k}|} \frac{M}{E ({\bf p})}\;
\left(\frac{m_W^2}{q^2-m_W^2}\right)^2
L^{\alpha \beta}_{\nu, \bar\nu}
\; W_{\alpha \beta}^{N}\,.
\end{equation}
When this cross section for a nucleon target moving with momentum $p^\mu$ in the rest frame of the nucleus of 
density $\rho(r)$ and weighted with the  spectral function $S_h (p^0, \bbox{p}, \rho(r))$, is summed over 
all the nucleons in the nucleus, it leads to Eq.~(\ref{cross_nuclear}). Similar equations are also obtained in 
Refs.~\cite{Petti,Petti1,Sargsian,Atti}. 
\begin{figure}
\includegraphics{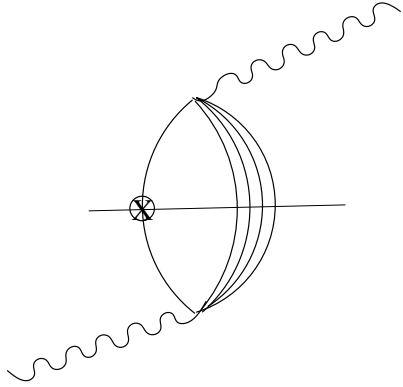}
\caption{Self-energy diagram of the W-boson in the nuclear medium. The imaginary part is calculated by cutting along the horizontal line and applying the Cutkosky rules when putting the particle on mass shell.}
\end{figure}

Now to evaluate $F^A_3$, we calculate the components $xy$
on both sides of Eq.~(\ref{conv_WA}). We get by taking $\bbox{q}$ along the $z$ axis:
\begin{equation}
W^{A}_{xy}= -\frac{i}{2M_A} q_z W^A_3\,,
\end{equation}
and for the right hand side we have for the nucleon moving with momentum $p$:
\begin{equation}	\label{W_N}
W^N_{xy}= \frac{p_x p_y}{M^2} W^N_{2} (p,q) +
\frac{i}{2M^2} W^N_{3}[p_z q_0 -p_0 q_z]\,.
\end{equation}
Since we have
\begin{equation}
q_0 W^A_{3}=F^A_3 (x)\,,
\end{equation}
\begin{equation}
\frac{p.q}{M} W^N_{3} (p,q) = F^N_3 (x_N)\,,
\end{equation}
with $x$ as defined in Eq.~(\ref{Bj_var}) and $x_N$ is the Bjorken
variable expressed in terms of the nucleon variables , $(p^0, {\bf p})$, in
the nucleus
\begin{equation}
x_N=\frac{Q^2}{2p.q}=\frac{Q^2}{2(p_0q_0-p_zq_z)},
\end{equation}
we obtain the expression for $F^A_3(x)$ in the Bjorken limit
\begin{eqnarray} 
F^A_{3} (x, Q^2)&=& 4 \int d^3 r \; \int \frac{d^3 p}{(2 \pi)^3} 
\; \frac{M}{E (\bbox{p})} \; \int^{\mu}_{- \infty} \; d p^0 S_h (p^0, \bbox{p}, \rho(r))
\nonumber\\
&&\times
\frac{x_N}{x}\left[\frac{p_0 q_z-p_z q_0}{Mq_z}
\right]F^N_{3}(x_N, Q^2)\,,
\end{eqnarray}
where the contribution of $W_2$ appearing in Eq.~(\ref{W_N}) vanishes
after momentum integration.
\begin{figure}[h]
\includegraphics{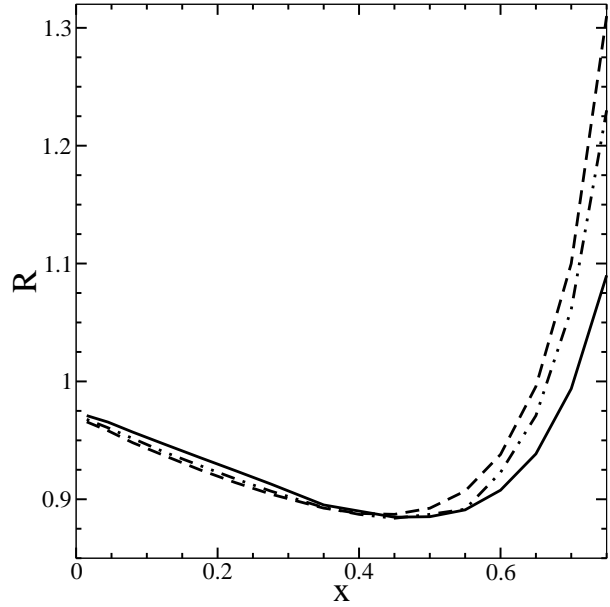}
\caption{Results for the ratio $R=\frac{F^A_{3} (x)}{A F^N_{3} (x)}$ at different $Q^2$ using MRST NNLO~\cite{MRST} parton distribution function.
 Solid line at $Q^2$=10$GeV^2$, dashed-dotted line at $Q^2$=100$GeV^2$ and dashed line at $Q^2$=1000$GeV^2$.}
\end{figure}
\begin{figure}[h]
\includegraphics{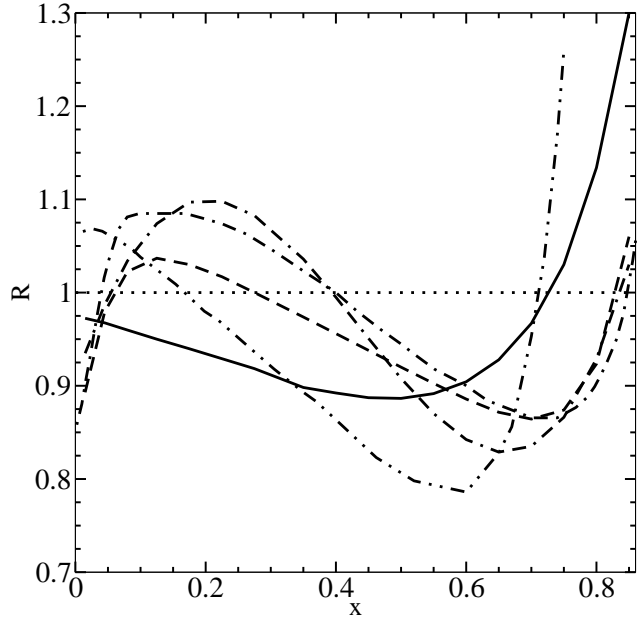}
\caption{Results for the ratio $R=\frac{F^A_{3} (x)}{A F^N_{3} (x)}$ at $Q^2=5$ GeV$^2$
by different authors. Solid line: this work using MRST2004 NNLO parton distribution function; double dashed-dotted line: Hirai et al.
\protect~\cite{Hirai}; short dashed line: NuTeV collaboration~\cite{Tzanov}, dashed-double dotted line the results of Kulagin~\cite{Kulagin} and the results of Kulagin and Petti~\cite{Petti1} shown by dashed-dotted line.}
\end{figure}
Defining $\gamma$ as
\begin{equation}	\label{gamma}
\gamma=\frac{q_z}{q^0}=
\left(1+\frac{4M^2x^2}{Q^2}\right)^{1/2}\,,
\end{equation}
we get
\begin{eqnarray} 	\label{finalF3}
F^A_3 (x, Q^2)&=& 4 \int d^3 r \; \int \frac{d^3 p}{(2 \pi)^3} 
\int^{\mu}_{- \infty} \; d p^0 S_h (p^0, \bbox{p}, \rho(r))
F(p, Q^2) F^N_3(x_N, Q^2),
\end{eqnarray}
where $F(p, Q^2)=\frac{M}{E({\bf p})}\left(\frac{p_0 \gamma-p_z}{(p_0-p_z\gamma) \gamma}\right)$.

This is our main equation which describes the modification on $F^A_3 (x, Q^2)$ due to nuclear medium effects and was earlier obtained in Ref.{\cite{Singh}}.

\section{Results and Discussion}
  We obtain the numerical results for $F^A_3(x, Q^2)$ using Eq.(\ref{finalF3}) and $S_h$ given by Eq.~(\ref{sh}). The density for $^{56}$Fe nucleus is taken to be a two Fermi parameter distribution i.e.
\begin{equation} \label{density}
\rho(r)=\frac{\rho_0}{1 + exp(\frac{r-c1}{c2})},
\end{equation}
 where the parameters c1=4.106fm and c2=0.519fm and are taken from Ref~\cite{Vries}.
While using the spectral function $S_h (\omega, {\bf p}, \rho(r))$ in Eq.(\ref{finalF3}), we ensure that it gives the correct binding energy for the iron nucleus by varying the free piece of the real part of the nucleon self energy $\Sigma^N$
in Eq.(\ref{sh}).

 This is done by calculating the average kinetic and total nucleon energy given by: 
\begin{eqnarray}
<T>= \frac{4}{A} \int d^3 r \;  \int \frac{d^3 p}{(2 \pi)^3} (E({\bf p})-M) 
\int^{\mu}_{- \infty} \; S_h (p^0, \bbox{p}, \rho(r)) 
\; d p^0\,,
\end{eqnarray}
\begin{eqnarray}
<E>= \frac{4}{A} \int d^3 r \;  \int \frac{d^3 p}{(2 \pi)^3}  
\int^{\mu}_{- \infty} \; S_h (p^0, \bbox{p}, \rho(r)) 
\; p^0 d p^0\,,
\end{eqnarray}
and the binding energy per nucleon given by~\cite{Marco}:
\begin{equation}
|E_A|=-\frac{1}{2}(<E-M>+\frac{A-1}{A-2}<T>)
\end{equation}
After the free parameter in $\Sigma^N$ is fixed, the expected values of the kinetic energy $<T>$ and the total nucleon energy $<E>$ are obtained as 30MeV and 48MeV for the case of $^{56}Fe$ which are in good agreement with other models and experiment (See e.g. Table II of Ref.\cite{Simula}). A similar agreement is also obtained with other nuclei. The structure function ${F^A_{3} (x, Q^2)}$ is numerically evaluated.
For $F^N_{3}(x_N,Q^2)$  we use the MRST2004~NNLO~parton distribution functions~\cite{MRST}. 
The $Q^2$ evolution of ${F^N_3(x, Q^2)}$ is assumed to be given by the $Q^2$ evolution of GLS integral $S(Q^2)$ given 
in Eq. (12). The strong coupling constant $\alpha_S(Q^2)$ is calculated using the variable flavour number evolution 
equation with $\Lambda=251MeV$ for $n_f=4$ and $\Lambda=178MeV$ for $n_f=5$~\cite{Larin} which gives $\alpha_S(M_Z)$=.1153.
 The constants $a(n_f)$ and $b(n_f)$ in Eq. (12) are taken from Larin and Vermarseren~\cite{Larin}. The target mass
corrections are incorporated using the prescription of Petti and Kulagin~\cite{Petti} following the work of Georgi and Politzer~\cite{Georgi}.
\begin{figure}
\includegraphics{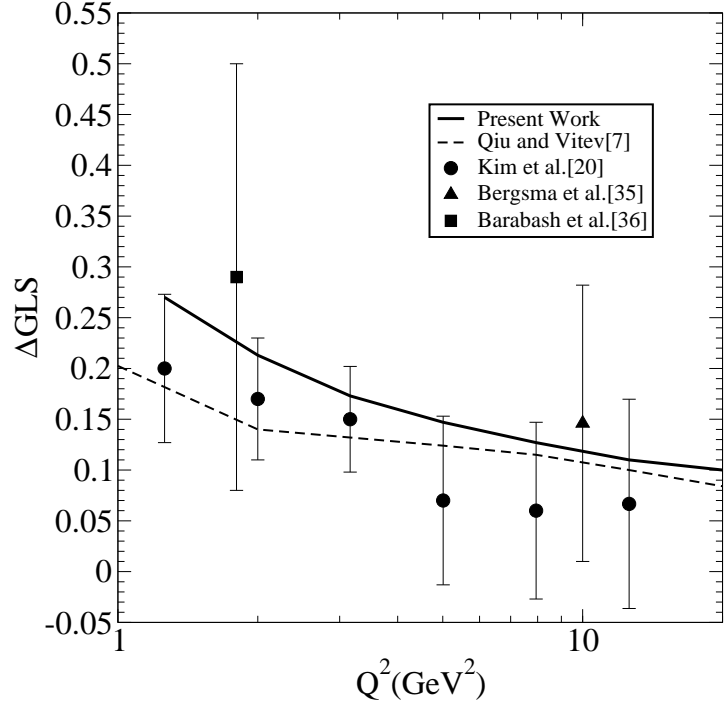}
\caption{$\Delta$GLS=$\frac{1}{3}(3-\int_0^1F^A_3(x, Q^2)dx)$ vs $Q^2$.}
\end{figure}
\begin{figure}
\includegraphics{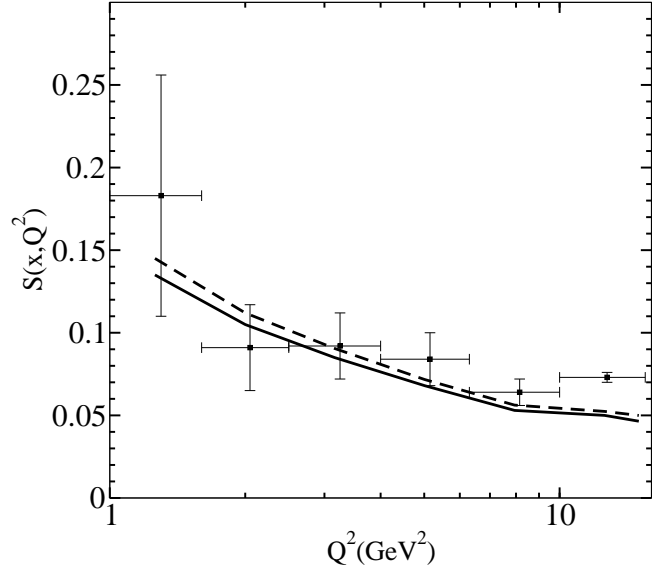}
\caption{The contribution to the GLS integral from the region x=0.5 to x=1 for different $Q^2$ using MRST NNLO\cite{MRST} parton distribution function(PDF) and its comparison from the results of Kim et al.\cite{Kim}. The dashed line is the result with the free case, solid line is the result with nuclear medium effects obtained in our model with spectral function including QCD and TMC corrections, and the solid square with error bars are the results of Kim et al.\cite{Kim}}
\end{figure}
\begin{figure}
\includegraphics{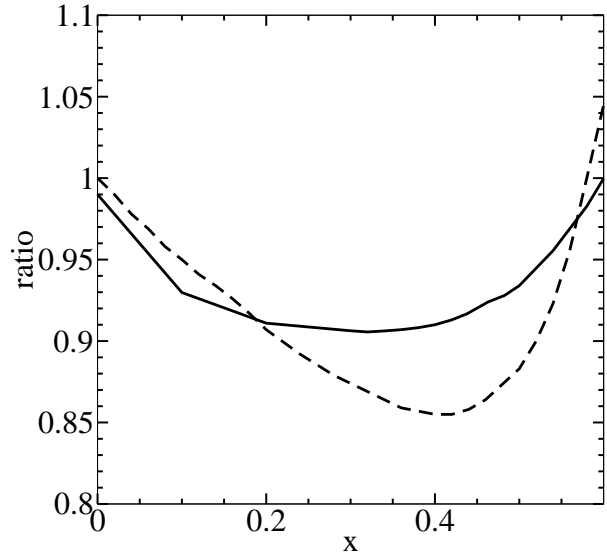}
\caption{The x dependence of the nucleus/nucleon ratio of the GLS integral from the nuclear effects to the free case i.e. the ratio $\frac{\int_{x_{min}}^1 F^A_3(x)dx}{\int_{x_{min}}^1  F^N_3(x)dx}$ at $Q^2=5GeV^2$ using MRST NNLO\cite{MRST} parton distribution function(PDF). The dashed line is the result of Kulagin~\cite{Kulagin}.}
\end{figure}
 
 In Fig.4, we show the results for the ratio R($x$,$Q^2$)=$\frac{F^A_3(x,Q^2)}{AF^N_3(x,Q^2)}$ as a function of $x$ for various values of $Q^2$. For $Q^2$=10$GeV^2$, we get a suppression for $x \le x_{min}$=0.7, beyond which we get an enhancement. This value of ${x}_{min}$ decreases with increasing $Q^2$. For $x > x_{min}$, the ratio increases very fast and becomes larger than unity as $x\rightarrow$ 1. This is mainly due to the  Fermi motion of the nucleons. In the region of 0.3$<x<$1, this behavior is very similar to that seen in the EMC effect with charged lepton and (anti)neutrino scattering from the nuclear targets in the analyses of the structure function $F^{A,l}_2(x, Q^2)$ and $F^{A, \nu(\bar\nu)}_2(x,Q^2)$.
  
 In the region of 0.1$<x<$0.3, we do not see any enhancement due to the anti-shadowing effect in $F^A_3(x, Q^2)$ as found in
$F^{A,l}_2(x, Q^2)$~\cite{Marco}. In the present calculation, this is due to the neglect of the mesonic contributions,  responsible for the enhancement in $F^{A,l}_2(x, Q^2)$. However, for $F^A_3(x, Q^2)$ these mesonic contributions are expected to be small because of  the vanishing of the dominant term due to pions. Even in the case of $F^{A, \nu(\bar\nu)}_2(x,Q^2)$, the recent\cite{Schienbein} estimates of nuclear medium effects in the anti-shadowing region, do not see any enhancement in this region, in disagreement with the other results available from earlier analyses~\cite{Hirai, Bodek1, Eskola}. This is not unexpected as these structure functions, at low x, probe different contributions of the parton distribution functions  in the nuclear medium. This suggests that the nuclear medium modification effects could be different in $F^{A,l}_2(x, Q^2)$, $F^{A,\nu(\bar\nu)}_2(x, Q^2)$ and $F^{A,\nu(\bar\nu)}_3(x, Q^2)$, and calls for a comprehensive study of nuclear medium modification effects on these structure functions specially in the region of small $x$.

In Fig.5, we compare our results for R($x$,$Q^2$) with the results of Tzanov et al.~\cite{Tzanov}, Kulagin and
Petti~\cite{Petti}, Kulagin~\cite{Kulagin} and Hirai et al.~\cite{Hirai}. While the work of Kulagin~\cite{Kulagin} and
Kulagin and Petti~\cite{Petti,Petti1} use a nuclear model to calculate the nuclear effects which shows a $Q^2$ dependence, the
work of Tzanov et al.~\cite{Tzanov} and Hirai et al.~\cite{Hirai} are phenomenological analyses, which assume the nuclear
effects to be independent of $Q^2$. In this figure, the results for R($x$,$Q^2$) for $Q^2=5GeV^2$, are presented. We find a
suppression in $F^A_3(x, Q^2)$ for $x<$0.7 and an enhancement thereafter,
 which are respectively smaller than the results of Ref.~\cite{Kulagin}, but are larger than the recent results of Kulagin
and Petti~\cite{Petti1}. It should be noted that
these latter results~\cite{Petti1} give  suppression in the region of 0.4$<x<$0.8 and enhancement for $x>$0.8, which
are smaller than the present results and the results obtained earlier in Ref.~\cite{Kulagin}. When compared with the results
of Tzanov et al.~\cite{Tzanov} and Hirai et al.~\cite{Hirai}, we find a smaller suppression in the region 0.5$<x<$0.7. 
In the region 0.7$<x<$0.8, we find an enhancement while they obtain a suppression. The results of the
phenomenological analyses shown in Fig.~5~\cite{Tzanov, Hirai} use the experimental data on ${F^{A,l}_2 (x, Q^2)}$ in charged
lepton scattering to estimate the nuclear effects. It can, therefore, be inferred that we find nuclear effects in the case of
${F^A_3 (x, Q^2)}$ different than obtained in the case of ${F^{A, l}_2} (x, Q^2)$. This is true for all $Q^2$ in our model. 

 The GLS sum rule (Eq.\ref{gls}) provides a benchmark to test various models used for the calculation of $F^A_3(x,Q^2)$. We find that in the $Q^2 \rightarrow \infty$ limit, the correction to the GLS sum rule in our model comes from the off shell modifications to the nucleon spectral function. It is easy to see this from Eq.(\ref{finalF3}). The nuclear GLS sum rule in the limit $Q^2\rightarrow \infty$ (i.e. $\gamma \rightarrow$1) is given by:
\begin{eqnarray} 	\label{finalgls}
S=\int_0^1F^A_3 (x)dx= 4 \int d^3 r \; \int \frac{d^3 p}{(2 \pi)^3} 
\int^{\mu}_{- \infty} \; d p^0 S_h (p^0, \bbox{p}, \rho(r))
\left(\frac{p_0-p_z}{E({\bf p})}
\right)F^N_3(x_N) dx_N~~
\end{eqnarray}

In the limit of noninteracting nucleons this trivially reproduces the GLS sum rule for free nucleons.

The $Q^2$ dependent nuclear effects in the GLS integral enter through the factor $\gamma$ and the $Q^2$ dependence of $F^N_3(x_N,Q^2)$ in Eq.(\ref{gamma}) when it is integrated over $x$. The $Q^2$ dependent nuclear corrections to the GLS sum rule are thus linked to the $Q^2$ dependent perturbative and non-perturbative QCD effects appearing in $F^N_3(x_N,Q^2)$.  In Fig.6, we show the $Q^2$ dependence of the nuclear effects of the GLS integral, where we plot $\Delta$GLS=$\frac{1}{3}(3-\int_0^1F^A_3(x, Q^2)dx)$ as a function of $Q^2$. The experimental results from CCFR collaborations\cite{Kim}, CHARM collaborations\cite{Bergsma} and IHEP-JINR collaborations\cite{Barabash} are also shown. The $Q^2$ behavior of $\Delta$GLS has been found to be in agreement with the present available experimental results. In this figure, we have also shown the theoretical results obtained by Qiu and Vitev~\cite{Qiu}. Our results are in agreement with the results of Qiu and Vitev~\cite{Qiu} for $Q^2>5GeV^2$ where theoretically the suppression is found to be larger than the experimental results. For $Q^2<5GeV^2$, we find a larger suppression compared to the central value of the experimental result and both theoretical values are within the experimental errors. The nuclear corrections to the GLS sum rule found by Kulagin and Petti~\cite{Petti1} are quite small due to the cancellation of nuclear shadowing, not included in our model, and off shell effects.

We show in Fig.7, the value of S($x$,$Q^2$)=$\int_x^1 dx F^A_{3} (x, Q^2)$ for $x$=0.5, with and without the nuclear effects and compare them with the experimental results of Kim et al.\cite{Kim}. The nuclear effects tend to reduce S($x$,$Q^2$) but the reductions are smaller than the experimental uncertainties. In Fig.8, we show the results for S($x$,$Q^2$)=$\int_x^1 dx F^A_{3} (x, Q^2)$ vs $x$ at $Q^2$=5$GeV^2$, where we have also shown the result obtained by  Kulagin~\cite{Kulagin}. We find that nuclear medium effects leading to a suppression in 0.2$<x<$0.55 region and an enhancement in $x>$0.55 region are smaller than obtained by Kulagin~\cite{Kulagin}.

To summarize our results, we have studied nuclear effects in the structure function $F^A_{3} (x, Q^2)$ in the iron nucleus using a many body theory to describe the spectral function of the nucleon in the nuclear medium for all $Q^2$. The nuclear effects are found to decrease the value of the structure function for $x \le x_{min}$=0.7 and increase at higher $x>x_{min}$. The parameter $x_{min}$ is found to be $Q^2$ dependent which decreases with $Q^2$. The results are compared with other theoretical analysis of Kulagin~\cite{Kulagin} and phenomenological analyses of Tzanov et al.~\cite{Tzanov} and Hirai et al.~\cite{Hirai}. The effect of the nuclear medium modifications on the Gross-Llewellyn Smith~\cite{GLS} sum rule and its $Q^2$-dependence has also been studied. In general nuclear medium effects decrease the value of GLS integral for all $Q^2$.
\section{Acknowledgment}
M. S. A. and S. K. S. wish
to acknowledge the financial support from the University of Valencia and Aligarh Muslim University under the academic exchange program. The authors thank E. Oset, University of Valencia for many useful discussions and encouragement throughout this work. We are also thankful to I. Ruiz Sumo, University of Valencia for useful discussions and rechecking our results. This work was partially supported by MEC contract FIS2006-03438, by the Generalitat Valenciana contract
ACOMP07/302  and by the EU Integrated
Infrastructure Initiative Hadron Physics Project contract
RII3-CT-2004-506078.

\end{document}